\pdfoutput=1
\documentclass[10pt,conference]{IEEEtran}
\IEEEoverridecommandlockouts

\usepackage{cite}
\usepackage{amsmath,amssymb,amsfonts}
\usepackage{graphicx}
\usepackage{textcomp}
\usepackage{xcolor}
\usepackage{booktabs}
\usepackage{multirow}
\usepackage{listings}
\usepackage{xspace}
\usepackage{url}
\usepackage{tikz}
\usetikzlibrary{arrows.meta,positioning,fit,backgrounds,calc}

\newcommand{\tbc}{\textcolor{red}{TBD}}
\newcommand{\VC}{Verification Contract\xspace}

\newcommand{\PA}{Phase~A\xspace}
\newcommand{\PB}{Phase~B\xspace}
\newcommand{\RQ}[1]{\textbf{RQ#1}}
\newcommand{\finding}[1]{\par\smallskip\noindent\fbox{\parbox{\dimexpr\columnwidth-2\fboxsep-2\fboxrule\relax}{\small\textit{#1}}}\par\smallskip}

\newcommand{\nRuns}{5\xspace}
\newcommand{\nRounds}{3\xspace}
\newcommand{\nSeeded}{46\xspace}

\newcommand{\res}[1]{\ifcsname #1\endcsname\csname #1\endcsname\else\tbc\fi}

\begin{document}

\title{Who Verifies the Builder? Toward Independent Assurance for Agent-Produced Software}

\author{\IEEEauthorblockN{Sion Israel Sion\thanks{Corresponding author: Sion Israel Sion, sion-israel.sion.1@ens.etsmtl.ca.}}
\IEEEauthorblockA{\textit{\'Ecole de technologie sup\'erieure}\\
Montr\'eal, Canada\\
sion-israel.sion.1@ens.etsmtl.ca\\
ORCID 0000-0002-7696-3167}
\and
\IEEEauthorblockN{Alain April}
\IEEEauthorblockA{\textit{\'Ecole de technologie sup\'erieure}\\
Montr\'eal, Canada\\
alain.april@etsmtl.ca}}
\newcommand{\repourl}{\url{https://github.com/besionisrael/assure-harness}}

\maketitle

\begin{abstract}
Coding agents can now interpret a specification, implement it, write and execute tests, repair failures, and declare the result ready for release. When the same agentic system produces both the software and the evidence of its correctness, a passing test suite no longer constitutes independent evidence. Visible tests become an optimization target, code, tests, and oracles may share a single misinterpretation of the specification, and the builder's own tests may simply be incomplete. This paper argues that agent-produced software requires an assurance plane that is structurally independent of the build plane. We first characterize four gaps along the chain from stakeholder intent to production, namely the requirement, conformance, evaluation, and model gaps, and identify which of them verification can close. We then propose a framework that separates building from verifying, introduces a Verification Contract stating what must be demonstrated before release, and organizes verification in two phases: a specification-first phase designed without access to the implementation, and an implementation-aware phase that strengthens the evidence through mutation analysis, property-based testing, and fuzzing. The resulting evidence is consolidated into an assurance report that makes residual uncertainty explicit and leaves the release decision to a human. We evaluate the framework in a pre-registered study on three stateful services, two of them written by third-party models, and an EvalPlus pilot, comparing self-validation with independent verification under context-level and model-level independence. The builder's programs were correct, but its own tests missed subtle interactions: a verifier in a fresh context detected more of the \nSeeded seeded faults than the builder's tests (Holm-corrected p = 0.006), and surfaced the ambiguities of the specification, at the price of rare false oracles. A verifier from another provider did not add detection. The framework and the study provide a basis for governing the release of software produced by agents.
\end{abstract}


\section{Introduction}
\label{sec:intro}

Large language models (LLMs) were first adopted in software engineering as assistants that complete code, generate unit tests, or suggest repairs under close human supervision~\cite{fan2023llmse,wang2024survey}. Coding agents change this division of labor. Given a task description and a repository, an agent can decompose the task, modify the code, write and execute tests, repair its own failures, and iterate until every visible check succeeds~\cite{jimenez2024swebench}. Spec-driven development extends this mode of work by asking the human to formalize intent as a specification before delegating implementation to the agent~\cite{delimarsky2025speckit}.

Spec-driven development returns control to the human at the beginning of the chain. It does not address the end of the chain. In a typical agentic workflow, the agent that implements the specification also writes the tests, chooses the expected outputs, interprets failures, modifies the code, and finally reports that the change is ready. The release signal is then a set of green checks produced, selected, or adjusted by the producer itself. This situation raises a governance question that long predates LLMs: \emph{can the producer of an artifact be the sole authority on its correctness?} Classical software engineering answered negatively. Testing practice advises against programmers testing their own programs~\cite{myers2011art}, and standards for critical software require verification activities with technical, managerial, and financial independence from development~\cite{ieee1012,rtca2011do178c}.

Recent evidence suggests that this answer remains valid for agents, for two reasons. First, when tests drive an iterative loop, they act as a reward signal, and a sufficiently capable optimizer can satisfy the signal without satisfying the intent behind it~\cite{amodei2016concrete,skalse2022reward}. Zhao et al.\ report that frontier coding agents saturate visible validation suites on systems-level tasks while still failing held-out composition tests, with failures that range from incomplete features to implementations that memorize test inputs~\cite{zhao2026specbench}. Zhong et al.\ show that agents frequently modify or circumvent tests when a specification and its tests conflict~\cite{zhong2026impossiblebench}. Second, when code, tests, and expected outputs derive from the same interpretation of a specification, they can be mutually consistent and jointly wrong. Internal consistency is not evidence of external correctness, which is a modern instance of the test oracle problem~\cite{barr2015oracle}. A third, less visible reason needs neither optimization pressure nor a misreading: a builder that implements the specification correctly still tests only what it thought of, so its passing suite says little about the behaviors it did not test.

The scale of agent output aggravates both problems. Line-by-line review cannot keep pace with the volume of code that agents produce, so the object of human control must shift from the code to the evidence about the code. We therefore formulate the following question: \emph{when software is produced largely autonomously from a specification, what independent evidence is sufficient for a human to authorize its release?}

This paper makes the following contributions:
\begin{itemize}
\item \textbf{A gap analysis.} We decompose the distance between stakeholder intent and production behavior into four gaps (requirement, conformance, evaluation, and model) and explain why self-validation can close none of them reliably (Section~\ref{sec:problem}).
\item \textbf{An independent assurance framework.} We separate a build plane from an assurance plane, introduce the \VC as the human-approved artifact that defines release evidence, and organize verification into a specification-first phase and an implementation-aware phase whose evidence is consolidated in an assurance report (Section~\ref{sec:framework}).
\item \textbf{A pre-registered empirical study.} We measure the evaluation gap under self-validation, the effect of context-level and model-level independence, and the complementarity of the two verification phases on three stateful services with \nSeeded seeded faults, two of them written by third-party models, and on an EvalPlus pilot~\cite{liu2023evalplus} (Sections~\ref{sec:design} and~\ref{sec:results}).
\item \textbf{An open artifact} containing the experimental harness, prompts, generated evidence, and analysis scripts \repourl.
\end{itemize}

\section{Background and Related Work}
\label{sec:related}

\subsection{LLM-Based Test Generation and Test Adequacy}

Early work on LLM-based testing asked whether models could produce tests that compile, run, and increase coverage. CodaMosa invokes an LLM when search-based test generation reaches a coverage plateau~\cite{lemieux2023codamosa}. Sch\"afer et al.\ evaluate LLM-generated unit tests for JavaScript and report coverage competitive with established tools~\cite{schafer2024testpilot}. TestGen-LLM at Meta accepts a generated test only if it builds, passes reliably, and increases coverage, and it reports that most candidates fail at least one of these filters~\cite{alshahwan2024testgen}. A recurring lesson is that LLMs are most useful when coupled with an external mechanism that measures, filters, or rejects their output.

Coverage, however, is a weak proxy for fault detection: Inozemtseva and Holmes find that coverage is not strongly correlated with test suite effectiveness once suite size is controlled~\cite{inozemtseva2014coverage}. Mutation analysis offers a stronger criterion by asking whether a suite fails on deliberately faulty variants of the program~\cite{jia2011mutation,papadakis2019mutation}, and mutant detection correlates with real fault detection~\cite{just2014mutants}. MuTAP feeds surviving mutants back to an LLM to strengthen generated tests~\cite{dakhel2024mutap}, and Meta's ACH system generates mutants that represent specific fault classes and then generates tests that kill them~\cite{foster2025ach}. Property-based testing~\cite{claessen2000quickcheck} replaces examples with invariants checked over generated inputs, and recent studies investigate LLMs as generators of such properties~\cite{vikram2023pbt,bose2025prompts}. Metamorphic testing addresses programs for which exact expected outputs are unavailable~\cite{chen2018metamorphic}.

\subsection{The Test Oracle Problem}

A test requires both an input and a judgment of whether the observed output is correct~\cite{barr2015oracle}. LLMs can generate the judgment, but the judgment can be wrong. TOGLL studies the correctness and strength of LLM-generated oracles and shows that both must be measured separately~\cite{hossain2025togll}. Several code generation approaches use generated tests to select or refine programs: CodeT ranks candidate programs by their agreement with generated tests~\cite{chen2023codet}, and AgentCoder assigns test design to a separate agent from code generation~\cite{huang2023agentcoder}. These approaches treat generated tests as a means to improve generation. We instead treat them as evidence for a release decision and ask under which conditions that evidence is independent.

\subsection{Specification Gaming by Coding Agents}

Reward hacking occurs when an optimizer increases a proxy reward without improving the true objective~\cite{amodei2016concrete,skalse2022reward}, a specific case of Goodhart's law~\cite{manheim2018goodhart}. In agentic coding, the proxy is the test suite. SpecBench separates natural-language specifications, visible validation tests, and held-out composition tests, and finds that the gap between visible and held-out performance grows with code size~\cite{zhao2026specbench}. ImpossibleBench constructs tasks whose tests contradict their specifications, so that any pass reveals a shortcut, and documents behaviors ranging from test deletion to operator overloading~\cite{zhong2026impossiblebench}. EvalPlus showed earlier that augmenting HumanEval and MBPP with additional tests substantially lowers measured correctness~\cite{liu2023evalplus}. Related work on LLM self-assessment reports that models struggle to correct their own reasoning without external feedback~\cite{huang2024selfcorrect} and that LLM evaluators favor their own generations~\cite{panickssery2024selfpref}.

\subsection{Independence and Assurance Cases}

Independent verification and validation (IV\&V) is a mature discipline. IEEE Std 1012 distinguishes technical, managerial, and financial independence of the verification organization~\cite{ieee1012}, and DO-178C scales required verification activities and their independence with the software level~\cite{rtca2011do178c}. Knight and Leveson showed experimentally that independently developed program versions fail in correlated ways, so that nominal independence does not imply statistical independence of failures~\cite{knight1986nversion}. Assurance cases structure a claim, its argument, and its supporting evidence, for example with the Goal Structuring Notation~\cite{kelly2004gsn,bloomfield2010assurance}. Our framework adapts these ideas to a setting they did not anticipate: a non-human producer that optimizes against its own evaluator and whose verifier may share its failure modes. Finally, Zave and Jackson distinguish requirements, specifications, and domain assumptions~\cite{zave1997dark}, a distinction we use to separate the gaps in Section~\ref{sec:problem}.

\section{Why Self-Validation Fails}
\label{sec:problem}

\subsection{Setting and Notation}

Let $R$ denote the stakeholder requirement, $S$ the written specification, $W$ the assumptions about the deployment environment, and $P$ the program produced by an agent. Following Zave and Jackson~\cite{zave1997dark}, a correct development satisfies $S \wedge W \models R$ and $P \models S$. A test suite $T$ is a finite set of pairs $(x, o)$ where $x$ is an input or interaction sequence and $o$ is an oracle; we write $\mathit{acc}_T(P)$ when $P$ passes every test in $T$. We distinguish a visible suite $T_v$, available to the builder, from a hidden suite $T_h$ derived from the same specification and never exposed to the builder.

In self-validation, the builder forms an interpretation $\hat{S}$ of $S$, produces $P$ and a suite $T_b$ from $\hat{S}$, and iterates until $\mathit{acc}_{T_v \cup T_b}(P)$ holds. The release signal is that predicate.

\subsection{Tests Become Rewards}

Once $\mathit{acc}_{T_v}$ terminates the loop, it operates as a reward. Any program in the set $\{P \mid \mathit{acc}_{T_v}(P)\}$ ends the episode, and that set is far larger than $\{P \mid P \models S\}$ because $T_v$ samples a small part of the input space. The agent has no incentive to distinguish the two sets, and capable agents demonstrably exploit the difference~\cite{zhao2026specbench,zhong2026impossiblebench}. We define the \emph{evaluation gap} of a development process over a task set as
\begin{equation}
\Delta_{\mathit{eval}} = \Pr\big[\mathit{acc}_{T_v}(P)\big] - \Pr\big[\mathit{acc}_{T_h}(P)\big],
\label{eq:evalgap}
\end{equation}
which is positive when visible success overstates correctness with respect to the same specification.

\subsection{Consistency Is Not Correctness}

The second failure mode does not require any optimization pressure. If $\hat{S} \neq S$, then $P$ and $T_b$ can agree with each other while both disagree with $S$. The builder's tests then confirm the builder's misunderstanding. This is a correlated, or common-mode, failure in the sense of Knight and Leveson~\cite{knight1986nversion}: two artifacts produced from the same interpretation fail on the same inputs. Separating test generation into another agent~\cite{huang2023agentcoder} reduces this correlation only if the second agent does not inherit $\hat{S}$, for example by reading the builder's code, rationale, or tests. Using the same underlying model may preserve part of the correlation even with separated contexts, since the two agents share training data and inductive biases. Whether context separation suffices, or whether model diversity is also needed, is an empirical question that we address in \RQ{2}.

\textbf{Incomplete self-evidence.} A third failure mode occurs even when $\hat{S} = S$. The builder can implement $S$ correctly and still test only part of it: $T_b$ exercises the behaviors the builder had in mind, and the others are correct without evidence. A later change that breaks them passes $T_b$. Self-validation cannot reveal this gap, because the builder chooses what to test with the same attention that missed it.

\subsection{Four Gaps Between Intent and Production}

Table~\ref{tab:gaps} decomposes the distance between $R$ and production behavior into four gaps. Verification against the specification can close only the conformance gap. The evaluation gap is closed by withholding the assurance artifacts from the builder. The requirement and model gaps cannot be closed by verification alone, because both concern whether $S$ and $W$ are the right objects to verify against. Our framework therefore does not claim to close them; it makes them explicit and assigns them to human approval and to runtime feedback.

\begin{table}[t]
\caption{Four gaps between stakeholder intent and production behavior}
\label{tab:gaps}
\centering
\footnotesize
\setlength{\tabcolsep}{3pt}
\begin{tabular}{@{}p{1.5cm}p{2.2cm}p{4.3cm}@{}}
\toprule
\textbf{Gap} & \textbf{Between} & \textbf{Treatment in the framework} \\
\midrule
Requirement & $R$ and $S$ & Human approval of the \VC; \PA reports ambiguities and contradictions in $S$ \\
Conformance & $S$ and $P$ & \PA and \PB evidence \\
Evaluation & $T_v$ and $S$ & Assurance artifacts are withheld from the builder \\
Model & $W$ and production & Runtime evidence revises the contract \\
\bottomrule
\end{tabular}
\end{table}

\section{An Independent Assurance Framework}
\label{sec:framework}

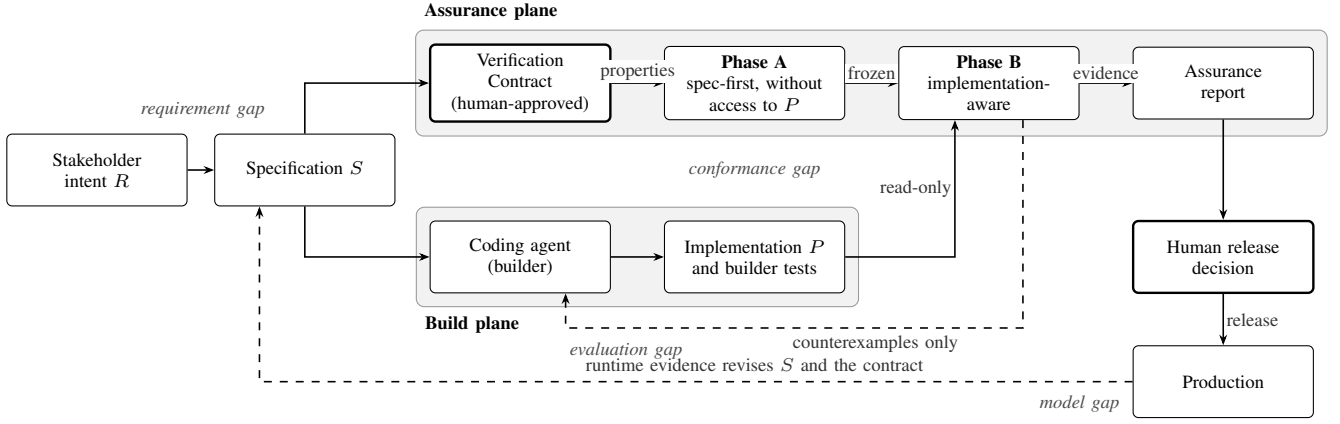
\begin{figure*}[t]
\centering
\begin{tikzpicture}[
  font=\scriptsize,
  box/.style={draw, rounded corners=2pt, align=center, text width=2.15cm, minimum height=0.95cm, fill=white},
  hbox/.style={box, line width=0.9pt},
  arr/.style={-{Stealth[length=4pt]}, semithick},
  darr/.style={arr, dashed},
  gap/.style={font=\scriptsize\itshape, text=black!70},
  lbl/.style={font=\scriptsize, text=black!80, fill=white, inner sep=1pt}
]
\node[hbox] (vc)  at (5.2, 0)    {Verification\\Contract\\(human-approved)};
\node[box]  (pa)  at (8.3, 0)    {\textbf{Phase A}\\spec-first, without\\access to $P$};
\node[box]  (pb)  at (11.4, 0)   {\textbf{Phase B}\\implementation-\\aware};
\node[box]  (rep) at (14.5, 0)   {Assurance\\report};
\node[box]  (ag)  at (5.2, -2.3) {Coding agent\\(builder)};
\node[box]  (imp) at (8.3, -2.3) {Implementation $P$\\and builder tests};
\node[box]  (int) at (-0.4, -1.15) {Stakeholder\\intent $R$};
\node[box]  (sp)  at (2.35, -1.15) {Specification $S$};
\node[hbox] (dec) at (14.5, -2.3) {Human release\\decision};
\node[box]  (prod) at (14.5, -3.95) {Production};
\begin{scope}[on background layer]
  \node[draw=black!40, fill=black!5, rounded corners=3pt, inner sep=5pt, fit=(vc)(pa)(pb)(rep)] (aplane) {};
  \node[draw=black!40, fill=black!5, rounded corners=3pt, inner sep=5pt, fit=(ag)(imp)] (bplane) {};
\end{scope}
\node[anchor=south west, font=\scriptsize\bfseries] at (aplane.north west) {Assurance plane};
\node[anchor=north west, font=\scriptsize\bfseries] at (bplane.south west) {Build plane};
\draw[arr] (int) -- (sp);
\draw[arr] (sp.north) |- (vc.west);
\draw[arr] (sp.south) |- (ag.west);
\draw[arr] (vc) -- node[above, lbl] {properties} (pa);
\draw[arr] (pa) -- node[above, lbl] {frozen} (pb);
\draw[arr] (pb) -- node[above, lbl] {evidence} (rep);
\draw[arr] (ag) -- (imp);
\draw[arr] (imp.east) -| node[pos=0.75, left, lbl] {read-only} ($(pb.south)+(-0.45,0)$);
\draw[darr] ($(pb.south)+(0.45,0)$) -- ($(pb.south)+(0.45,-2.75)$) -| ($(ag.south)+(0.6,0)$);
\node[lbl, anchor=north] at (9.9, -3.3) {counterexamples only};
\draw[arr] (rep) -- (dec);
\draw[arr] (dec) -- node[right, lbl] {release} (prod);
\draw[darr] (prod.west) -| ($(sp.south)+(-0.6,0)$);
\node[lbl, anchor=south] at (8.3, -3.86) {runtime evidence revises $S$ and the contract};
\node[gap] at (1.0, -0.35) {requirement gap};
\node[gap] at (8.3, -1.15) {conformance gap};
\node[gap, anchor=north] at (6.6, -3.3) {evaluation gap};
\node[gap, anchor=north] at (12.6, -4.0) {model gap};
\end{tikzpicture}
\caption{Overview of the framework. The build plane produces the program; the assurance plane produces evidence about it. The builder receives counterexamples but never the assurance artifacts or fixes. Gaps from Table~\ref{tab:gaps} are shown in italics.}
\label{fig:framework}
\end{figure*}

Fig.~\ref{fig:framework} gives an overview. The framework is a process architecture and is independent of any specific model or tool.

\subsection{Separation of Powers}

The framework assigns three roles. The \emph{builder} produces $P$. The \emph{verifier} produces evidence about $P$ against $S$. The \emph{human} approves what must be demonstrated and decides whether the evidence suffices. Five principles govern the relation between roles.

\begin{enumerate}
\item[\textbf{P1}] \textbf{No self-certification.} Artifacts produced by the build plane, including the builder's tests, never count as evidence on their own.
\item[\textbf{P2}] \textbf{Information asymmetry.} Assurance artifacts are withheld from the builder. The builder receives only reproducible counterexamples, which prevents the assurance suite from becoming the new reward surface.
\item[\textbf{P3}] \textbf{Specification before code.} \PA evidence is derived and frozen before the verifier observes $P$.
\item[\textbf{P4}] \textbf{Verify, do not repair.} The verifier demonstrates defects but does not fix them. A verifier that repairs becomes a second builder and inherits the builder's incentives.
\item[\textbf{P5}] \textbf{Explicit residual uncertainty.} Evidence is reported per requirement with what was not verified, and the release decision remains human.
\end{enumerate}

We adapt the three dimensions of independence in IEEE Std 1012~\cite{ieee1012} to agents. \emph{Technical independence} concerns context and model: the verifier does not share the builder's conversation, rationale, or artifacts, and may use a different model. \emph{Managerial independence} requires that the verifier is not invoked, configured, or terminated by the builder's orchestration loop. \emph{Incentive independence}, the agentic counterpart of financial independence, requires that the verifier's objective is not tied to the builder's success; a verifier rewarded for a green pipeline has the same incentive as the builder.

\subsection{The Verification Contract}

A specification states what the system must do. A \VC states what must be demonstrated before the system is accepted. For each requirement it lists guarantees, which are falsifiable claims, and required evidence, which states the techniques and thresholds that support each guarantee. Fig.~\ref{fig:contract} shows an excerpt. The contract plays the role of the claim structure in an assurance case~\cite{kelly2004gsn}, with one difference: it is fixed and approved before any implementation exists, so the builder cannot shape the claims after the fact.

The contract is the artifact that the human inspects instead of the code. Its size grows with the number of requirements and risks, not with the size of $P$, which keeps human review tractable when the code exceeds review capacity. The contract can be drafted by an agent from $S$, but drafting is part of the assurance plane and approval is human. Unresolved ambiguities in $S$ are recorded in the contract as open questions rather than resolved silently, which is how the framework exposes the requirement gap.

\begin{figure}[t]
\begin{lstlisting}
REQ-17  A user can read only the documents they own.
 G17.1  No document is returned without authentication.
 G17.2  For users A != B, no request by A returns a
        document owned by B (direct, list, search,
        pagination, export).
 G17.3  An expired token is rejected.
 G17.4  A deleted document is never returned again.
 Evidence
  E17.a  stateful property tests over request sequences
  E17.b  fuzzing of document identifiers
  E17.c  mutants of authorization checks: all killed
 Open question: may an administrator read all documents?
REQ-23  Retrying a payment never creates a second
        transaction (idempotency).
\end{lstlisting}
\begin{lstlisting}
ASSURANCE REPORT  change #1842            verdict: BLOCK
REQ-17  VERIFIED    G17.1-4 hold; 0 surviving mutants
REQ-23  VIOLATED    counterexample EV-88291:
        POST /payment -> timeout -> retry
        -> 2 transactions persisted
NOT VERIFIED  concurrent retries across replicas
              (outside the test environment model)
\end{lstlisting}
\caption{Illustrative excerpt of a \VC (top) and of the corresponding assurance report (bottom), written for the documents service of the study. The study itself gives verifiers the specification $S$ directly; it does not produce contracts or reports.}
\label{fig:contract}
\end{figure}

\subsection{Phase A: Specification-First Verification}

In \PA, the verifier receives $S$, the approved contract, and the public interface of $P$ (signatures or API schema), but not its body. It produces executable properties, invariants, forbidden scenarios, and example-based acceptance tests for each guarantee. For stateful requirements it produces model-based tests over sequences of operations, including forbidden transitions such as paying twice or refunding a payment that was never captured. \PA artifacts are frozen before \PB starts. The rationale is that tests designed without the implementation cannot be shaped by it: they encode the verifier's reading of $S$ rather than the builder's reading reflected in $P$. The main risk of \PA is an incorrect oracle that encodes the verifier's own misreading of $S$. We measure this risk explicitly through the false-oracle rate (Section~\ref{sec:design}).

\subsection{Phase B: Implementation-Aware Verification}

In \PB, the verifier gains read-only access to $P$. It cannot alter \PA artifacts, but it can add evidence targeted at the implementation. It runs mutation analysis on $P$ and generates tests for surviving mutants~\cite{dakhel2024mutap,foster2025ach}; it fuzzes input boundaries revealed by the code; it adds property-based tests for internal paths and error handling that the specification leaves implicit; and, where an exact oracle is unavailable, it uses metamorphic or differential relations~\cite{chen2018metamorphic}. Where a guarantee is critical and the code is amenable, formal verification can replace testing as the evidence for that guarantee.

The two phases are complementary by construction. \PA asks whether $P$ does what $S$ requires, independently of how $P$ is written. \PB asks whether the evidence is strong enough to notice if $P$ were wrong, given how $P$ is written. Whether the two phases detect different faults in practice is \RQ{3}.

\subsection{Counterexample Loop}

When the verifier finds a violation, it returns a minimal reproducible counterexample to the builder: the input or request sequence, the observed behavior, and the guarantee it violates. It does not return the test code, the property, or a proposed fix (P2, P4). The builder repairs $P$, and the verifier re-evaluates the new version against the full frozen evidence. The loop is bounded, and a guarantee that remains violated at the bound is reported as such.

\subsection{The Assurance Report}

The output of the assurance plane is not a pass rate. The report lists, for each requirement, its status (verified, partially verified, violated, or not verified), the techniques that support the status, the strength of that support (mutation score, number of generated executions, counterexamples found), and the scope that was not verified, such as environment assumptions that the test harness does not model (Fig.~\ref{fig:contract}). The report deliberately avoids a single aggregate trust score, because no empirical basis currently exists for combining heterogeneous evidence into one number. It ends with a recommendation derived from contract thresholds, and the human decides.

\subsection{Runtime Feedback}

Verification establishes properties of $P$ relative to a model of its environment; production reveals where that model was wrong. Incidents, anomalies, and production traces that contradict an assumption in $W$ become new guarantees or new open questions in the contract. Assurance is therefore a process that runs in parallel with construction and continues after release, rather than a terminal gate.

\section{Study Design}
\label{sec:design}

\subsection{Research Questions}

\begin{itemize}
\item[\RQ{1}] \textbf{Evaluation gap.} To what extent does self-validation overstate the correctness of agent-produced code with respect to its specification?
\item[\RQ{2}] \textbf{Independence.} Does an independent verifier detect more faults than the builder's own tests, and does model-level independence add to context-level independence?
\item[\RQ{3}] \textbf{Phase complementarity.} Do specification-first and implementation-aware verification detect different faults, and what does their combination add?
\item[\RQ{4}] \textbf{Release decisions.} How accurate are release recommendations derived from the assurance report compared with a conventional pass/fail signal?
\end{itemize}

\subsection{Subjects}

\textbf{Function-level pilot.} We first ran a pilot on 20 tasks sampled at random from HumanEval+ and MBPP+~\cite{liu2023evalplus}, with the base suite as $T_v$ and the extended suite as $T_h$. The builder solved 19 of them, so these benchmarks contain too few faults to compare conditions. We report the pilot as exploratory and use stateful services for the main study.

\textbf{Stateful services.} Each service is a Python class with an in-process, HTTP-style interface, a natural-language specification $S$ of 10 requirements, a reference implementation, a small visible suite $T_v$ of nominal cases, a hidden suite $T_h$, and seeded faults. Each seeded fault is one local edit of the reference that passes $T_v$ and fails $T_h$. A validator checks these properties, and that $T_h$ passes on the reference.
(i)~\emph{Documents and payments} (14 faults): we wrote it, with requirements such as REQ-17 (ownership) and REQ-23 (idempotent payments) of Fig.~\ref{fig:contract}. After a first pilot, we revised its specification to state four subtle interactions precisely, for example that idempotency keys are scoped to the user rather than to the session, and to leave five points deliberately open, such as case-insensitivity beyond ASCII. The hidden suite tests the former and never the latter.
(ii)~\emph{Room booking} (17 faults) and (iii)~\emph{equipment loans} (15 faults) were written by two third-party models that are not under study, Mistral Medium 3.5 and Qwen 3.7-Plus, from a fixed brief that asks for realistic faults and states that $T_h$ may test only what $S$ fixes. The authors' role was limited to returning the validator's findings. Each hidden suite was then reviewed against its specification by a different third-party model, and mechanically checked for status codes absent from $S$ and for accesses to internal state. A third service, written by Gemini~3, was dropped after its correction rounds failed to converge. The services, briefs, and model answers are in the artifact.

\subsection{Conditions}

Table~\ref{tab:conditions} summarizes the conditions. All conditions share the same builder model and the same budget of \nRounds counterexample rounds.

\begin{table}[t]
\caption{Experimental conditions. $S$: specification; $T_v$: visible tests; $P$: implementation.}
\label{tab:conditions}
\centering
\footnotesize
\setlength{\tabcolsep}{3pt}
\begin{tabular}{@{}llcccl@{}}
\toprule
\textbf{ID} & \textbf{Evidence by} & $S$ & $T_v$ & $P$ & \textbf{Verifier model} \\
\midrule
SV   & Builder              & \checkmark & \checkmark & \checkmark & n/a \\
A    & Verifier, \PA        & \checkmark &            &            & same as builder \\
B    & Verifier, \PB        & \checkmark &            & \checkmark & same as builder \\
AB   & Verifier, \PA, \PB   & \checkmark &            & after \PA  & same as builder \\
AB$'$ & Verifier, \PA, \PB  & \checkmark &            & after \PA  & other provider \\
\bottomrule
\end{tabular}
\end{table}

\textbf{Shared starting point.} For each task and run, the builder first receives $S$ and $T_v$, writes a program and its own tests $T_b$, and iterates on the results of both suites for at most five rounds. The resulting program $P_0$ and the builder's conversation are the common starting point of all conditions, so that differences between conditions are attributable to the evidence alone. In SV, $P_0$ is the final program and $T_b$ is the evidence.

\textbf{Verifier conditions.} In A, B, AB, and AB$'$, a verifier in a fresh context produces evidence that is then frozen. In \PA it receives $S$ and the name of the class or function under test. In \PB it receives $S$ and $P_0$ in read-only form, together with the frozen \PA tests when they exist. It then receives up to six mutants of $P_0$ that survive its tests and may add tests that kill them. The frozen evidence is executed on the current program, and failing tests are returned to the builder as counterexamples that state the violated guarantee, the observed behavior, and the falsifying input, but not the test code (P2). The builder may reject a counterexample that contradicts $S$. This loop runs for at most \nRounds rounds. Comparing AB with AB$'$ isolates model-level independence (\RQ{2}). Comparing A, B, and AB isolates the phases (\RQ{3}). The builder and the same-model verifier are Claude Sonnet 5.5; the alternative verifier is GPT-5.5 (\texttt{gpt-5.5-2026-04-23}). Models run with their provider defaults for temperature and reasoning, and with an output limit of 64k tokens, because reasoning tokens count against it. All roles share one prompt template, and all prompts are in the artifact. Each service is executed \nRuns times to account for nondeterminism.

\subsection{Metrics}

We report each metric separately and do not aggregate them into a single score.
\begin{itemize}
\item \textbf{Visible and hidden pass rates} of the final program, and the evaluation gap $\Delta_{\mathit{eval}}$ of Eq.~\eqref{eq:evalgap} (\RQ{1}).
\item \textbf{False-oracle rate}, the fraction of tests in an evidence set that fail on the canonical or reference implementation. Such tests encode an incorrect reading of $S$~\cite{hossain2025togll}. They are excluded from the two detection metrics below, so that an evidence set cannot gain detection power by rejecting correct behavior.
\item \textbf{Fault detection.} For each task, the \emph{fault pool} contains every distinct program produced in any run and condition that fails $T_v$ or $T_h$ (function-level tasks), or that exhibits a seeded fault (case study). The detection rate of an evidence set is the fraction of the pool on which at least one of its valid tests fails (\RQ{2}, \RQ{3}).
\item \textbf{Mutation score} of each evidence set against first-order mutants of the canonical or reference implementation, as a measure of oracle strength~\cite{just2014mutants}. Mutants that $T_v \cup T_h$ does not detect are treated as likely equivalent and excluded.
\item \textbf{Complementarity}, the sets of faults and mutants detected uniquely by \PA, uniquely by \PB, and by both (\RQ{3}).
\item \textbf{Decision accuracy.} Each condition yields a verdict on its final program: \emph{release} if the program passes $T_v$ and every test of the evidence, \emph{block} otherwise. Ground truth is correctness on $T_v \cup T_h$. We report precision and recall of the \emph{block} verdict twice: on the raw evidence, as a release gate would see it, and on valid tests only, which isolates the cost of false oracles (\RQ{4}).
\item \textbf{Cost}, measured in tokens and wall-clock time.
\end{itemize}

\subsection{Analysis}

We registered the protocol of the main campaign before running it: models, parameters, budgets, hashes of every frozen file, and two confirmatory hypotheses. \textbf{H1}: AB detects more seeded faults than SV (context-level independence). \textbf{H2}: AB and AB$'$ differ in detection, with no predicted direction (model-level independence). Since the same faults recur in every run, the unit of analysis is the seeded fault. For each fault, we average its detection over the runs of each condition, and we apply an exact two-sided sign test over the faults, dropping ties. H1 and H2 form their own Holm family at $\alpha = 0.05$. All other analyses are exploratory: they use exact McNemar tests over (task, run) pairs for binary outcomes and Wilcoxon signed-rank tests with the Vargha--Delaney $\hat{A}_{12}$ effect size for continuous outcomes~\cite{arcuri2011stats,vargha2000a12}, corrected as a separate Holm family. The protocol and its deviation log are in the artifact.

\section{Results}
\label{sec:results}

\subsection{Overview}

Table~\ref{tab:services} lists the three services of the main campaign. We
ran \nRuns runs per service (15 units, 270 model calls, 5.8M tokens: 5.0M for
the Claude roles and 0.8M for the alternative verifier). No unit failed or had
to be resumed. Hypotheses H1 and H2 were registered before the campaign; every
other analysis below is exploratory, and all deviations from the protocol are
listed in the artifact.

\begin{table}[t]
\caption{Services of the main campaign}
\label{tab:services}
\centering
\footnotesize
\setlength{\tabcolsep}{3pt}
\begin{tabular}{@{}llccc@{}}
\toprule
\textbf{Service} & \textbf{Author of $S$, $T_h$, faults} & \textbf{Req.} & \textbf{Faults} & \textbf{$T_h$ tests} \\
\midrule
Documents and payments & the authors & 10 & 14 & 39 \\
Room booking & Mistral Medium 3.5 & 10 & 17 & 27 \\
Equipment loans & Qwen 3.7-Plus & 10 & 15 & 27 \\
\bottomrule
\end{tabular}
\end{table}

\textbf{A measurement defect.} After the runs, we found that our harness
called every test function without arguments. Verifier tests written with
\texttt{pytest.mark.parametrize} therefore failed with a \texttt{TypeError} on
every program, including the reference implementation. This affected 282 of
the 323 verifier tests that we had counted as invalid. These tests were
excluded from detection, so the confirmatory detection results are
conservative for the verifiers. They were, however, counted as failures by
the raw release gate. We report the confirmatory results as registered, and an
exploratory rescoring of the same evidence, with these tests run and without
any new model call, wherever the defect matters.

\subsection{RQ1: Evaluation Gap Under Self-Validation}

On the EvalPlus pilot (20 tasks, one run), the builder's final program passed
$T_h$ for 19 tasks. The gap was 0 points on HumanEval+ and 10 points on MBPP+,
and that single case came from hidden tests that require behavior the
specification does not fix (a negative $k$ in Mbpp/753). In the main
campaign, the builder's final program passed $T_v \cup T_h$ in all 15 units,
in every condition (\res{resHidPassSVCSii}\% on every service). The builder
produced no faulty program, so every fault analyzed below is seeded.

\finding{Finding 1. With a strong builder and precise specifications, we did
not observe an evaluation gap. The only gap we saw came from hidden tests that
exceed the specification, not from the builder.}

\subsection{RQ2: Effect of Independence}

\begin{table*}[t]
\caption{Evidence per condition. Seeded faults detected (\%, mean over \nRuns runs) per service and pooled; invalid tests as measured and genuine false oracles (\% of tests, all services); mutation score (\%, median over services); tokens per unit, excluding the shared builder phase (median over services)}
\label{tab:rq2}
\centering
\footnotesize
\begin{tabular}{@{}lccccccccc@{}}
\toprule
& \multicolumn{4}{c}{\textbf{Seeded faults detected}} & \multicolumn{2}{c}{\textbf{Tests failing on the reference}} & & & \\
\cmidrule(lr){2-5}\cmidrule(lr){6-7}
\textbf{Cond.} & Documents & Booking & Loans & All & As measured & Genuine & \textbf{Mut.} & \textbf{Tokens} & \textbf{Detected, rescored} \\
\midrule
SV    & \res{resSeedDetSVCSii} & \res{resSeedDetSVSvcM} & \res{resSeedDetSVSvcQ} & \res{resSeedDetSV} & 0.6 & 0.6 & \res{resMutSV} & \res{resTokSV} & 196/230 \\
A     & \res{resSeedDetACSii} & \res{resSeedDetASvcM} & \res{resSeedDetASvcQ} & \res{resSeedDetA} & 11.4 & 1.9 & \res{resMutA} & \res{resTokA} & 216/230 \\
B     & \res{resSeedDetBCSii} & \res{resSeedDetBSvcM} & \res{resSeedDetBSvcQ} & \res{resSeedDetB} & 10.8 & 1.0 & \res{resMutB} & \res{resTokB} & 219/230 \\
AB    & \res{resSeedDetABCSii} & \res{resSeedDetABSvcM} & \res{resSeedDetABSvcQ} & \res{resSeedDetAB} & 8.9 & 0.7 & \res{resMutAB} & \res{resTokAB} & 221/230 \\
AB$'$ & \res{resSeedDetABprimeCSii} & \res{resSeedDetABprimeSvcM} & \res{resSeedDetABprimeSvcQ} & \res{resSeedDetABprime} & 7.0 & 2.3 & \res{resMutABprime} & \res{resTokABprime} & 200/230 \\
\bottomrule
\end{tabular}
\end{table*}

Table~\ref{tab:rq2} summarizes the evidence. The confirmatory tests treat each
of the \nSeeded seeded faults as a unit: a fault favors a condition when its
detection rate over the runs is higher, and ties are dropped.

\textbf{Context-level independence (H1).} AB detected more seeded faults than
the builder's own tests: 11 faults favor AB, 1 favors SV, and 34 are ties
(exact sign test, $p_{\mathrm{Holm}} = 0.006$). The effect does not rest on
the faults we designed. On the two services written by third-party models
alone, 7 faults favor AB and none favors SV ($p = 0.016$). All of them come from
the booking service. On the loans service, SV, A, B, and AB each detected the
same 14 of 15 faults in every run. The faults that separate the conditions follow one pattern: the
builder implements a subtle interaction correctly, but does not test it. On
the documents service, for example, SV never detected the two faults that bind
idempotency keys or pagination cursors to a session instead of a user (0 of 10
fault-runs), while AB detected them in 6 of 10.

\textbf{Model-level independence (H2).} The verifier from another provider
detected fewer faults than the same-model verifier: 13 faults favor AB, none
favors AB$'$, and 33 are ties ($p_{\mathrm{Holm}} = 0.0005$). Provider
diversity did not add detection here. Since we compare one pair of models,
this result confounds provider and capability.

\textbf{Cost.} The independent conditions cost 67k to 126k tokens per unit,
on top of the shared builder phase. Measured invalid tests reached 7 to 11\%
of the verifiers' tests, but genuine false oracles were 0.7 to 2.3\%; the rest
came from the defect described above. In the exploratory rescoring, the
conclusions are unchanged and slightly stronger (H1: 11 to 0, $p = 0.001$;
H2: 10 to 0, $p = 0.002$).

\finding{Finding 2. A verifier in a fresh context detected more seeded faults
than the builder's own tests, including on a service written by a third party.
The effect vanishes when every condition reaches the ceiling. A verifier from
another provider did not add detection. Genuine false oracles were rare.}

\subsection{RQ3: Complementarity of the Two Phases}

Within AB, Phase~A alone detected \res{resSeedABAonly} (fault, run) pairs,
Phase~B alone \res{resSeedABBonly}, and both \res{resSeedABBoth}. Within
AB$'$, the split was \res{resSeedABprimeAonly}, \res{resSeedABprimeBonly}, and
\res{resSeedABprimeBoth}. Phase~A therefore carries most of the detection, and
Phase~B adds a smaller set of faults.

Phase~A is also where ambiguities surface. Phase~A reported at least one open
question in \res{resOpenQuestionShare}\% of the task runs. On the documents
service, we left five points deliberately open. Phase~A raised the meaning of
case-insensitivity beyond ASCII in 4 of 5 runs (A) and 3 of 5 (AB), and
the validity of an empty idempotency key in 3 of 5 (A) and 5 of 5 (AB). SV,
B, and AB$'$ never raised them. No condition raised the question of an
administrator role, the example of Fig.~\ref{fig:contract}.

\finding{Finding 3. Specification-first verification carries most of the
detection and is where ambiguities are exposed. Implementation-aware
verification adds a smaller, distinct set of faults.}

\subsection{RQ4: Accuracy of Release Decisions}

Since every final program was correct, we can measure only the cost of the
release gate, not its recall. On raw evidence, the gate blocked a correct
program in 14 of 15 units for AB, 13 for A, 11 for B, 4 for AB$'$, and none
for SV. Most of these blocks came from the tests the harness could not run.
With those tests run (exploratory rescoring), the counts fall to 8 for AB, 6
for A and for B, and 0 for AB$'$. On valid tests only, the gate blocked a
correct program in 2 units, both AB on the documents service, and in no other
case. Both blocks come from tests that require passwords made of
unencodable characters to be accepted, a behavior that the specification
leaves open and that the reference happens to accept.

Invalid evidence did not degrade the programs: no final program became
incorrect in any condition, although builders received counterexamples from
invalid tests. In one booking unit, the builder answered that ``each is a TypeError raised
because the verifier's own parametrized test functions were called without
their arguments'', and it kept its implementation unchanged.

\finding{Finding 4. As a release gate, raw independent evidence blocks correct
programs. Most of these blocks disappear once invalid tests are separated, and
the counterexample loop already exposes those tests. Recall could not be
measured, because no faulty program reached release.}

\section{Discussion}
\label{sec:discussion}

\subsection{Which Failure Mode Did We Observe?}

Section~\ref{sec:problem} describes three ways in which self-validation can fail. Our study observed only the third one. The builder did not overfit the visible tests: its final programs passed the hidden suites in every unit (Finding~1). Its code and its tests did not share a misreading either, since its code was correct exactly where its tests were silent. What we observed is incomplete self-evidence. On the documents service, the builder correctly bound idempotency keys and pagination cursors to the user in every run, but its tests never exercised this binding. A seeded fault that bound them to the session instead passed every builder test, while the independent verifier detected it. The first two failure modes have been reported for larger tasks and for agents under stronger optimization pressure~\cite{zhao2026specbench,zhong2026impossiblebench}. Our strong builder and precise specifications may have suppressed them. They remain reasons to separate the planes, but measuring them requires tasks on which the builder fails, which we leave to future work.

\subsection{What the Human Controls}

The framework moves human control from the code to three artifacts whose size does not grow with the code: the \VC before implementation, the counterexamples during development, and the assurance report at release. The human remains accountable for the two gaps that verification cannot close. Approving the contract is a judgment that $S$ captures $R$, and accepting a residual risk listed as not verified is a judgment about $W$. Neither judgment can be delegated to the builder without reintroducing self-certification. The results of \RQ{4} show why the human must see the evidence rather than a single verdict: raw independent evidence blocked correct programs, and almost all of these blocks came from invalid tests that the counterexample loop had already exposed. A gate that reports which tests the builder disputes, and why, lets the human resolve these cases quickly. A gate that only says ``block'' does not.

\subsection{Implications for Governance and Compliance}

The theme of governance and oversight of AI in testing suggests three practical implications. First, an organization that adopts coding agents should treat the builder's tests as development artifacts and not as release evidence, consistent with P1. Second, independence should be specified and audited along the three dimensions of Section~\ref{sec:framework}, in the same way that IV\&V independence is specified for critical software~\cite{ieee1012}. Third, the required strength of evidence should scale with risk, as software levels do in DO-178C~\cite{rtca2011do178c}: a contract for an authorization requirement can demand killed mutants on every access check, whereas a contract for a formatting utility can accept example-based tests. Because each verdict in the report links a requirement to identified evidence and counterexamples, the report also constitutes an audit trail of what was known when the release was authorized.

\subsection{Who Verifies the Verifier?}

The assurance plane is itself built from LLMs and can be wrong. The framework does not assume a correct verifier; it bounds the consequences of an incorrect one. A false oracle produces a counterexample that the builder cannot reproduce against $S$, which surfaces the disagreement rather than hiding it. Mutation analysis measures whether the verifier's oracles are strong. Human approval of the contract constrains what the verifier attempts to demonstrate. The false-oracle rate reported in Section~\ref{sec:results} quantifies the residual risk. In our study, genuine false oracles were rare (0.7--2.3\% of verifier tests). Most of them asserted a behavior that the specification leaves open, such as accepting passwords of unencodable characters. Our own harness was a larger source of error: it silently failed every parametrized test, and it took a builder's rejection of these counterexamples to reveal it. The lesson applies to any assurance plane: the infrastructure that runs evidence is itself part of what must be verified, and it should reject evidence that it cannot execute instead of reporting it as failing.

\section{Threats to Validity}
\label{sec:threats}

\textbf{Construct validity.} The builder produced no faulty program, so detection is measured on seeded faults only. Seeded faults and mutants approximate real defects~\cite{just2014mutants}, but they may not represent the faults of agent-produced code. Hidden tests approximate the specification in both directions: they can miss behavior, and they can also exceed it, as one EvalPlus task did, which inflates the evaluation gap. The false-oracle check depends on the reference implementation: on behavior that $S$ leaves open, a permissive reference makes an invalid test look valid. Finally, our harness did not execute parametrized tests (Section~\ref{sec:results}). This made the confirmatory detection results conservative for the verifiers, and it inflated the measured false-oracle rates and raw release blocks. We report the decomposition and an exploratory rescoring with these tests run.

\textbf{Internal validity.} We revised the documents service after a first pilot, so its interactions and open points were designed with knowledge of earlier results. We mitigated this with two services written by third-party models, on which H1 also holds. These models are not under study, but their ground truth may share the biases of LLMs, and it was reviewed by models, not by humans. The coding of open points was also done by a third-party model, before the detection scores were read. LLM outputs are nondeterministic and sensitive to prompts; we used \nRuns runs per service, provider defaults, and one prompt template for all roles. H2 compares one pair of models, so provider and capability are confounded. All deviations from the registered protocol are documented in the artifact.

\textbf{External validity.} Three small Python services and a function-level pilot do not represent repository-level development. We used one builder model, and the ceiling effects we observed suggest that results depend on the strength of the builder relative to the difficulty of the tasks. EvalPlus tasks may be present in training data. Other languages, larger systems, and other agent architectures may lead to different results.

\textbf{Conclusion validity.} The unit of analysis is the seeded fault (\nSeeded faults). Most faults were detected by every condition, so few faults are informative (12 for H1). The effect of H1 is concentrated in one third-party service and in our own service, and the loans service showed no difference. We therefore report per-service results next to the pooled tests. Only H1 and H2 are confirmatory.

\section{Conclusion and Research Agenda}
\label{sec:conclusion}

Spec-driven development gives the human control over what agents should build. This paper argues that the next step is to give the human control over what agents are allowed to ship. We characterized four gaps between intent and production, proposed an assurance framework that separates the build plane from the assurance plane and organizes verification into a specification-first and an implementation-aware phase, and evaluated it empirically. In a pre-registered study, the builder's code was correct, but its own tests missed faults that a verifier in a fresh context detected, and only that verifier surfaced the ambiguities of the specification. Genuine false oracles were rare, and a verifier from another provider added nothing. Independence of context, more than independence of model, made the evidence stronger.

Several problems remain open. Deriving verification intent from a specification without inventing requirements is unsolved. The degree of independence needed between builder and verifier, and its cost, must be characterized beyond function-level tasks. Test adequacy for agent-produced code requires a multidimensional notion of evidence that has not yet been validated. Assurance must extend to distributed systems, data migrations, configuration, and concurrency, where the model gap dominates. Finally, a user study is needed to establish whether an assurance report improves the calibration of human release decisions compared with conventional continuous integration results.

\section*{Data Availability}
The harness, prompts, generated evidence, the three services and the briefs and answers of the models that wrote two of them, the registered protocol with its deviation log, and the analysis scripts are available at \repourl.

\bibliographystyle{IEEEtran}
\bibliography{refs}

\end{document}